\documentclass[twoside,reqno]{lt6-proc}
\usepackage{epsfig,cite}
\usepackage{amssymb,amsmath}
\usepackage{times}
\begin{document}
\sloppy \raggedbottom
\setcounter{page}{1}

\newpage
\setcounter{figure}{0}
\setcounter{equation}{0}
\setcounter{footnote}{0}
\setcounter{table}{0}
\setcounter{section}{0}

\newtheorem{theo}{Theorem}
\newtheorem{defi}[theo]{Definition}
\newtheorem{lemm}[theo]{Lemma}
\newtheorem{coro}[theo]{Corollary}
\newtheorem{rema}[theo]{Remark}
\newtheorem{prop}[theo]{Proposition}

\def\bea{\begin{eqnarray}}
\def\eea{\end{eqnarray}}
\def\nn{\nonumber}
\def\beq{\begin{equation}}
\def\eeq{\end{equation}}
\def\nn{\nonumber}
\def\ds{\displaystyle}
\def\C{\mathbb{C}}
\def\Z{\mathbb{Z}}
\def\ra{\rangle}
\def\la{\langle}
\def\lb{[\![}
\def\rb{]\!]}
\def\l{\ldots}
\def\t{\theta}
\def\q{{\bar q}}
\def\L{{\bar L}}


\title{Lie superalgebraic framework for generalization of quantum statistics}

\runningheads{N.I. Stoilova and J. Van der Jeugt}{Generalization of Quantum Statistics}

\begin{start}


\author{N.I. Stoilova}{1,2},
\coauthor{J. Van der Jeugt}{1}

\address{Department of Applied Mathematics and Computer Science,\\ University of Ghent,
Krijgslaan 281-S9, B-9000 Gent, Belgium\\
E-mails~: Neli.Stoilova@UGent.be, Joris.VanderJeugt@UGent.be}{1}
\address{Institute for Nuclear Research and Nuclear Energy,\\ Boul. Tsarigradsko Chaussee
72, 1784 Sofia, Bulgaria}{2}


\begin{Abstract}
Para-Bose and para-Fermi statistics are known to be associated with 
representations of the  Lie (super)algebras of class $B$. We develop 
a framework for the generalization of quantum statistics
based on the  Lie superalgebras $A(m|n)$, $B(m|n)$, $C(n)$ and $D(m|n)$.
\end{Abstract}
\end{start}


\section{Introduction}

It has been known for more than 50 years that generalizations of ordinary
Bose and Fermi quantum statistics are possible if one abandons the requirement for 
the commutator or anticommutator of two fields to be a $c$-number. The commutation (resp.
anticommutation) relations between the Bose (resp. Fermi) creation and annihilation
operators (CAOs) can be replaced by a weaker system of triple relations for the so-called 
para-Bose operators~\cite{Green}
\begin{eqnarray}
&& [\{ B_{ j}^{\xi}, B_{ k}^{\eta}\} , B_{l}^{\epsilon}]=
(\epsilon -\xi) \delta_{jl} B_{k}^{\xi} +  (\epsilon -\eta)
\delta_{kl}B_{j}^{\eta}, \nn \\ 
&& \xi, \eta, \epsilon =\pm; \ j,k,l=1,\ldots,n
\label{para-Bose} 
\end{eqnarray}
and para-Fermi operators\cite{Green}
\begin{eqnarray}
&& [[F_{ j}^{\xi}, F_{ k}^{\eta}], F_{l}^{\epsilon}]=\frac 1 2
(\epsilon -\eta)^2
\delta_{kl} F_{j}^{\xi} -\frac 1 2  (\epsilon -\xi)^2
\delta_{jl}F_{k}^{\eta}, \nn \\
&& \qquad \xi, \eta, \epsilon =\pm\hbox{ or }\pm 1;\quad j,k,l=1,\ldots,n. \label{para-Fermi}
\end{eqnarray}
It was shown  by Kamefuchi and Takahashi~\cite{K}, and by Ryan and Sudarshan~\cite{S},
that the Lie algebra generated by the $2n$ elementss $F_i^\xi$  subject to the relations~(\ref{para-Fermi}) is 
the Lie algebra $so(2n+1)\equiv B_n$. 
Similarly Ganchev and Palev~\cite{Ganchev} discovered a new connection, namely between para-Bose
statistics and the orthosymplectic Lie superalgebra (LS) $osp(1|2n)\equiv B(0|n)$~\cite{Kac}.
The LS generated by $2n$ odd elements 
 $B_i^\xi$, subject to the 
relations~(\ref{para-Bose}) is 
$osp(1|2n)\equiv B(0|n)$~\cite{Kac}.
Therefore para-statistics can be associated with representations of  the Lie (super)algebras
of class $B$.
Alternative types of generalized quantum statistics 
in the framework of other classes of simple Lie algebras or superalgebras have
been considered in particular by Palev~\cite{Palev1}-\cite{sl(1|n)}. 
 Furthermore, a complete classification of 
all the classes of generalized quantum statistics for the classical 
Lie algebras $A_n$, $B_n$, $C_n$ and $D_n$, by means of their algebraic relations,
was given in~\cite{GQS}.
In the present paper we make a similar classification for the basic
classical Lie superalgebras.

\section{Preliminaries, definition and classification method}

Let $G$ be a basic classical Lie superalgebra. $G$ has a $\Z_2$-grading $G=G_{\bar 0}\oplus G_{\bar 1}$;
an element $x$ of $G_{\bar 0}$ is an even element ($\deg(x)=0$), 
an element $y$ of $G_{\bar 1}$ is an odd element ($\deg(y)=1$). 
The Lie superalgebra bracket is denoted by $\lb x,y \rb$. In the universal enveloping
algebra of $G$, this stands for 
\[
\lb x,y \rb = x y - (-1)^{\deg(x)\deg(y)} y x,
\]
if $x$ and $y$ are homogeneous. So the bracket can be a commutator or an anti-commutator.

A generalized quantum statistics associated with 
$G$ is determined by  $N$ creation operators $x_i^+$ and $N$ annihilation
operators $x_i^-$. Inspired by the para-statistics, Palev's statistics and~\cite{GQS},
these $2N$ operators should generate the Lie superalgebra $G$, subject to certain triple
relations. Let $G_{+1}$ and $G_{-1}$ be the subspaces of $G$ spanned by the CAOs:
\begin{equation}
G_{+1} = \hbox{span} \{x^+_i;\ i=1\ldots,N\},\qquad
G_{-1} = \hbox{span} \{x^-_i;\ i=1\ldots,N\}.
\end{equation}
We do not require that these subspaces are homogeneous.
Putting $G_{\pm 2}=\lb G_{\pm 1},G_{\pm 1}\rb$ and $G_0=\lb G_{+1},G_{-1}\rb$, 
the condition that $G$ is generated by the $2N$ elements subject to triple relations
only, leads to the requirement that $G= G_{-2} \oplus G_{-1} \oplus G_0 \oplus G_{+1} \oplus G_{+2}$,
and this must be a $\Z$-grading of $G$.
Since these subspaces are not necessarily homogeneous, this $\Z$-grading is in general
not consistent with the $\Z_2$-grading.

We impose two further requirements: first of all,
the generating elements $x_i^\pm$ must be root vectors of $G$.
Secondly, $\omega(x_i^+)=x_i^-$, where $\omega$ is the standard antilinear anti-involutive
mapping of $G$ (in terms of root vectors $e_\alpha$, $\omega$ satisfies
$\omega(e_\alpha)=e_{-\alpha}$).
This leads to the following definition:

\begin{defi}
Let $G$ be a basic classical Lie superalgebra, with antilinear anti-involutive mapping $\omega$.
A set of $2N$ root vectors $x^\pm_i$ ($i=1,\ldots,N$) is called a set of
creation and annihilation operators for $G$ if:
\begin{itemize}
\item $\omega(x^\pm_i)=x^\mp_i$,
\item $G= G_{-2} \oplus G_{-1} \oplus G_0 \oplus G_{+1} \oplus G_{+2}$ is
a $\Z$-grading of $G$, with $G_{\pm 1}= \hbox{span}\{x^\pm_i,\ i=1\ldots,N\}$
and $G_{j+k}=\lb G_j,G_k \rb$.
\end{itemize}
The algebraic relations ${\cal R}$ satisfied by the operators $x_i^\pm$
are the relations of a generalized quantum statistics (GQS) associated with $G$.
\end{defi}

A consequence of this definition is that the algebraic relations 
${\cal R}$ consist of quadratic and triple relations only. Another 
consequence is that 
$G_0$  is a subalgebra of $G$
spanned by root vectors of $G$, i.e. $G_0$ is a regular subalgebra 
of $G$.
By the adjoint action, the remaining $G_i$'s are $G_0$-modules.
Thus the following technique  can be used in order to obtain a complete classification
of all GQS associated with $G$:
\begin{enumerate}
\item
Determine all regular subalgebras $G_0$ of $G$. 
\item
For each regular subalgebra $G_0$, determine the decomposition of $G$
into simple $G_0$-modules $g_k$ ($k=1,2,\ldots$).
\item
Investigate whether there exists a $\Z$-grading of $G$ of the form
\begin{equation}
G=G_{-2} \oplus G_{-1} \oplus G_0 \oplus G_{+1} \oplus G_{+2}, 
\label{5grading}
\end{equation}
where
each $G_i$ is either directly a module $g_k$ or else a sum of
such modules $g_1\oplus g_2\oplus \cdots$, such that
$\omega (G_{+i})=G_{-i}$.
\end{enumerate}

If the $\Z$-grading is of the form~(\ref{5grading}) with $G_{\pm 2}\ne 0$, we shall say
that it has {\em length}~5; if $G_{+2}=0$ (then $G_{-2}=0$, but $G_{\pm1}\ne0$), then the
$\Z$-grading is of length~3.

In the following section we shall give a summary of the classification
process for the basic classical Lie superalgebras 
$A(m|n)$, $B(m|n)$, $B(0|n)$, $D(m|n)$ and $C(n)$. 
For more details on the classification techniques, see~\cite{CLS}.

\section{Classification}

\subsection{The Lie superalgebra $A(m|n)$}
Let $G$ be the special linear Lie superalgebra $A(m|n)\equiv sl(m+1|n+1)$, consisting of traceless
$(m+n+2)\times (m+n+2)$ matrices.  The root vectors of $G$ are known to be the
elements $e_{jk}$ ($j\ne k=1,\ldots,m+n+2$), where $e_{jk}$ is a matrix with
zeros everywhere except a 1 on the intersection of row~$j$ and column~$k$. The 
$\Z_2$-grading is such that $\deg(e_{jk})=\theta_{jk}=
\theta_j +\theta_k$, where 
\begin{eqnarray}
&&\theta_j= \left\{ \begin{array}{lll}
 {0} & \hbox{if} & j=1,\cdots ,m+1 \\ 
 {1} & \hbox{if} & j=m+2,\cdots ,m+n+2.
 \end{array}\right.
\end{eqnarray}
The root corresponding to $e_{jk}$ ($j,k=1,\ldots,m+1$) is given by $\epsilon_j-\epsilon_k$;
for $e_{m+1+j,m+1+k}$ ($j,k=1,\ldots,n+1$) it is $\delta_j-\delta_k$;
and for $e_{j,m+1+k}$, resp.\ $e_{m+1+k,j}$, ($j=1,\ldots,m+1$, $k=1,\ldots,n+1$) it is $\epsilon_j-\delta_k$,
resp.\ $\delta_k-\epsilon_j$.
The anti-involution is such that $\omega(e_{jk})=e_{kj}$.

In order  to find regular subalgebras of $G=A(m|n)$, one should delete nodes from the Dynkin diagrams of
$A(m|n)$ (first the ordinary,
and then the extended). 

\noindent
{\bf Step 1.} Delete node $i$ from the distinguished 
Dynkin diagram. Then  $A(m|n)=G_{-1}\oplus G_0\oplus G_{+1}$,
with
$G_0=sl(i)\oplus sl(m+1-i|n+1)$ for $i=1,\ldots,m+1$ and 
$G_0=sl(m+1|i-m-1)\oplus sl(n+m+2-i)$ for $i=m+2,\ldots,m+n+1$;
$ G_{-1}={\rm span}\{ e_{kl};\ k=1,\ldots ,i,\ l=i+1,\ldots ,m+n+2\};$
$ G_{+1}={\rm span}\{ e_{lk};\ k=1,\ldots ,i,\ l=i+1,\ldots ,m+n+2\}$ and 
$N=i(m+n+2-i)$.

For $i=1$, $N=m+n+1$. Putting
$
a_j^-=e_{1,j+1}, \quad a_j^+=e_{j+1,1},\qquad j=1,\cdots ,m+n+1, 
$
the relations ${\cal R}$ are:
\begin{eqnarray}
&&\lb a_j^+,a_k^+\rb =\lb a_j^-,a_k^-\rb=0,\nn\\
 &&\lb \lb a_j^+,a_k^-\rb ,a_l^+\rb =(-1)^{\t_{j+1}}\delta_{jk}a_l^++\delta_{kl}a_j^+ , \label{A1}
\\
&&\lb \lb a_j^+,a_k^-\rb ,a_l^-\rb =-(-1)^{\t_{j+1}}\delta_{jk}a_l^--
(-1)^{\t_{j+1,k+1}\t_{l+1}}\delta_{jl}a_k^- . \nn
\end{eqnarray}
For $m=0$, these are the relations of $A$-superstatistics~\cite{Palev5}, \cite{sl(1|n)}. 
Also for general $m$ and $n$, these relations have been considered in another context~\cite{sl(m|n)}.

For $i=2$, $N=2(m+n)$. One puts
\begin{eqnarray}
&& a_{-,j}^-=e_{1,j+2}, \qquad a_{+,j}^-=e_{2,j+2},\qquad j=1,\ldots ,m+n,\nn
\\
&&a_{-,j}^+=e_{j+2,1}, \qquad a_{+,j}^+=e_{j+2,2},\qquad j=1,\ldots ,m+n. \nn
\end{eqnarray}
Then the corresponding relations read ($\xi, \eta, \epsilon =\pm$; $j,k,l=1,\ldots,m+n$):
\begin{eqnarray}
  &&\lb a_{\xi j}^+,a_{\eta k}^+\rb =\lb a_{\xi j}^-,a_{\eta k}^-\rb =0, \nn\\
&& \lb a_{\xi j}^+,a_{-\xi k}^-\rb =0, \qquad  
 \lb a_{-j}^+,a_{- k}^-\rb =\lb a_{+j}^+,a_{+k}^-\rb , \qquad j\neq k, \label{Adouble} \\
&& \lb a_{+j}^+,a_{- j}^-\rb =\lb a_{+k}^+,a_{- k}^-\rb, \qquad\hbox{ for } \t_{j}=\t_{k}, \nn\\
&& \lb a_{-j}^+,a_{+ j}^-\rb =\lb a_{-k}^+,a_{+ k}^-\rb, \qquad\hbox{ for } \t_{j}=\t_{k}, \nn\\
 &&\lb\lb a_{\xi j}^+,a_{\eta k}^-\rb ,a_{\epsilon l}^+\rb=
 (-1)^{\deg(a_{\xi j}^+)\deg(a_{\eta k}^-)+\delta_{\xi,-\eta}\t_{12}
 \deg(a_{\epsilon l}^+)}
 \delta_{\eta\epsilon}\delta_{jk}
 a_{\xi l}^+\nn \\
&& \hskip 2.8cm +\delta_{\xi\eta}\delta_{kl}a_{\epsilon j}^+ , \nn
\\
&&\lb \lb a_{\xi j}^+,a_{\eta k}^-\rb ,a_{\epsilon l}^-\rb =-
(-1)^{\deg(a_{\xi j}^+)\deg(a_{\eta k}^-)}
\delta_{\xi \epsilon}\delta_{jk}
 a_{\eta l}^-\nn \\
 && \hskip 2.8cm -
 (-1)^{\t_{j+2,k+2}\deg(a_{\epsilon l}^-)}
 \delta_{\xi\eta}\delta_{jl}a_{\epsilon k}^- . \nn
\end{eqnarray}

\noindent {\bf Step 2.} 
Delete node $i$ and $j$ from the distinguished Dynkin diagram.
We have $G_0=H+sl(i)\oplus sl(j-i)\oplus sl(m+1-j|n+1)$ for
$1\leq i<j\leq m+1$, $G_0=H+sl(i)\oplus sl(m+1-i|j-m-1)\oplus sl(m+n+2-j)$ for
$1\leq i\leq m+1$, $m+2\leq j\leq m+n+1$ and 
$G_0=H+sl(m+1|i-m-1)\oplus sl(j-i)\oplus sl(m+n+2-j)$ for $m+2\leq i<j\leq m+n+1$. 
There are six simple $G_0$-modules. All the possible combinations 
of these modules give rise to gradings of length~5.
There are  three different ways in which these $G_0$-modules
can be combined. To characterize these three cases, it is sufficient
to give only $G_{-1}$:
\begin{eqnarray}
G_{-1}&=&\hbox{span}\{ e_{kl},e_{lp};\ k=1,\ldots ,i,
\ p=j+1, \ldots, m+n+2,\nn\\
&& l=i+1,\ldots ,j\},\;\;
\hbox{with }N=(j-i)(m+n+2-j+i); \label{A21}\\
G_{-1}&=&\hbox{span}\{ e_{kl},e_{pk};\ k=1,\ldots ,i,
\ p=j+1, \ldots, m+n+2\}, \nn \\
&& l=i+1,\ldots ,j,\; \hbox{with }N=i(m+n+2-i); \label{A22}\\
G_{-1}&=&\hbox{span}\{ e_{kl},e_{lp};\ k=1,\ldots ,i,
\  l=j+1, \ldots, m+n+2\}, \nn \\
&& p=i+1,\ldots ,j, \; \hbox{with }N=j(m+n+2-j).\label{A23} 
\end{eqnarray}
For  $j-i=1$ one can label the CAOs as follows:
$a_k^-=e_{k,i+1}, \; a_k^+=e_{i+1,k}, \; k=1,\ldots ,i;\;\;
a_k^-=e_{i+1,k+1}, \;\; a_k^+=e_{k+1,i+1}, \;\; k=i+1,\ldots ,m+n+1.
$
Using 
\begin{equation}
\langle k\rangle= \left\{ \begin{array}{lll}
 {0} & \hbox{if} & k=1,\ldots ,i, \\ 
 {1} & \hbox{if} & k=i+1,\ldots ,m+n+1,
 \end{array}\right.
\label{twokinds} 
\end{equation}
the quadratic and triple relations now read:
\begin{eqnarray}
 &&\lb a_k^+,a_l^+\rb =\lb a_k^-,a_l^-\rb =0, \; k,l=1,\ldots, i \ \hbox{or}\
 k,l =i+1, \ldots ,m+n+1,\nn\\
 && \lb a_k^-,a_l^+\rb =\lb a_k^+,a_l^-\rb =0, \; k=1,\ldots , i, \ l=i+1,\ldots, m+n+1,\nn \\
 &&\lb \lb a_k^+,a_l^-\rb ,a_p^+\rb =(-1)^{\la l\ra +\la p \ra +\la k \ra 
 \t_{k+1,i+1}}\delta_{kl}a_p^+\nn \\
 && \qquad \qquad \qquad \ + (-1)^{\la l\ra +\la p \ra +(1-\la l \ra )\t_{l,i+1}(\t_{lk}+\t_{k,i+1})} 
 \delta_{lp}a_k^+ ,\nn\\
 &&
 \qquad \qquad \qquad  \ \  k,l=1,\cdots, i,\ {\rm or} \ k,l=i+1,\ldots , m+n+1, \nn \\
&&\lb \lb a_k^+,a_l^-\rb ,a_p^-\rb =
-(-1)^{\la l\ra 
+\la p \ra + \deg(a_k^+)[\la k \ra \t_{k+1,l+1}+(1-\la l \ra)\t_{l,i+1}]}
\delta_{kp}a_l^-\nn\\
&&
\hskip 2.6cm -(-1)^{\la l\ra +\la p \ra +\la k \ra \t_{k+1,i+1}}\delta_{kl}a_p^-,
  \;  \nn\\
&& \hskip 2.5cm k,l=1,\cdots, i,\ {\rm or} \ k,l=i+1,\ldots , m+n+1, \nn \\
&&\lb \lb a_k^{\xi},a_l^{\xi}\rb ,a_p^{-\xi}\rb =
-(-1)^{{\frac{1} {2}}\t_{p,i+1}[(1+\xi)\t_{l+1,i+1}+
(1-\xi)\t_{k,l+1}]}\delta_{kp}a_l^{\xi}\nn \\
&& \hskip 2.6cm
+(-1)^{{\frac{1} {2}}(1+\xi)\t_{l+1,i+1}(\t_{k,i+1}+\t_{k,l+1})}
\delta_{lp}a_k^{\xi}, \nn\\
&&\hskip 2.5cm k=1,\ldots, i, \ l=i+1,\ldots ,m+n+1 ,\nn \\
&&\lb \lb a_k^{\xi},a_l^{\xi}\rb ,a_p^{\xi}\rb =0 ,\qquad 
\xi=\pm;\ k,l,p=1,\ldots,m+n+1.\label{A21R}  
\end{eqnarray}

\smallskip \noindent {\bf Step 3.} 
If we delete three or more nodes from the distinguished Dynkin diagram, the resulting
$\Z$-gradings of $A(m|n)$ are no longer of the required form.

\smallskip \noindent {\bf Step 4.} 
If we delete node $i$ from the extended distinguished Dynkin diagram,
the remaining diagram is again (a non-distinguished Dynkin diagram) of type $A(m|n)$, so $G_0=G$,
and there are no CAOs.

\smallskip\noindent {\bf Step 5.} 
Delete node $i$ and $j$ ($i<j$) from the extended distinguished Dynkin diagram.
Then
$A(m|n)=G_{-1}\oplus G_0\oplus G_{+1}$ with 
$G_0=H+sl(m|n+1)$ or $H+sl(m+1|n)$ when the nodes are adjacent,
and
$G_0=H+sl(k|l)\oplus sl(p|q)$
with $k+p=m+1$ and $l+q=n+1$ when the nodes are nonadjacent. 
\[
G_{-1}=\hbox{span}\{ e_{kl};\ k=i+1\ldots ,j,\ l\neq i+1,\ldots ,j\}
\]
and
$N=(j-i)(n+m+2-j+i)$.

\smallskip\noindent {\bf Step 6.} 
Delete nodes $i$, $j$ and $k$ from the extended distinguished 
Dynkin diagram ($i<j<k$).
For three adjacent nodes $G_0=H+sl(m-1|n+1)$, $H+sl(m|n)$ or $H+sl(m+1|n-1)$.
For two adjacent and one nonadjacent nodes  
$G_0=H+sl(l|p)\oplus sl(q|r)$ with $l+q=m$, $p+r=n+1$ or
$l+q=m+1$, $p+r=n$. 
If all three nodes are nonadjacent  
$G_0=H+sl(l|p)\oplus sl(q|r)\oplus sl(s|t)$ with $l+q+s=m+1$, 
$p+r+t=n+1$. One or two of these three Lie superalgebras is 
$sl(r|0)=sl(0|r)=sl(r)$.
There are three different ways in which the corresponding $G_0$-modules
can be combined. We give here only $G_{-1}$:
\begin{eqnarray}
G_{-1}&=&\hbox{span}\{ e_{ps},e_{sq};\ p=1,\ldots ,i,k+1,\ldots ,n+m+2, \nn\\
&& 
s=i+1,\ldots ,j,\ q=j+1, \ldots, k\},\nn\\
&& \hbox{with }N=(j-i)(n+m+2-j+i); \nn\\
G_{-1}&=&\hbox{span}\{ e_{ps},e_{qp};\ p=1,\ldots ,i,k+1,\ldots ,n+m+2,\nn\\
&&
\ s=i+1,\ldots ,j,\ q=j+1, \ldots, k\}, \nn\\
&& \hbox{with }N=(k-i)(n+m+2+i-k); \nn \\
G_{-1}&=&\hbox{span}\{ e_{pq},e_{qs};\ p=1,\ldots ,i,k+1,\ldots ,n+m+2,\nn \\
&&
\ s=i+1,\ldots ,j,\ q=j+1, \ldots, k\}, \nn\\
&& \hbox{with }N=(k-j)(n+m+2+j-k).\nn 
\end{eqnarray}

\smallskip\noindent {\bf Step 7.} 
If we delete four or more nodes from the extended distinguished 
Dynkin diagram  the 
$\Z$-grading of $A(m|n)$ satisfies no longer the required properties.

\smallskip\noindent {\bf Step 8.} 
Next,  one should repeat the process for all nondistinguished Dynkin
diagrams of $G$ and their extensions. 
The only new result corresponds to Step~6 deleting three nonadjacent nodes from
the extended Dynkin diagram. 
We have
$G_0=H+sl(l|p)\oplus sl(q|r)\oplus sl(s|t)$ with $l+q+s=m+1$, 
$p+r+t=n+1$ and in some cases none of the three algebras is 
$sl(r|0)=sl(0|r)=sl(r)$.

\subsection{The Lie superalgebras $B(m|n)$}

We summarize the classification process for the Lie superalgebras $B(m|n)$ giving for all nonisomorphic GQS 
the subalgebra $G_0$ 
(each $G_0$ contains the complete Cartan subalgebra $H$, so we only list the remaining part of 
$G_0=H+\cdots$); the length $\ell$ of the $\Z$-grading and  the number $N$ of annihilation operators:

\vskip 0.5cm
\begin{tabular}{l|l|l}
\hline 
&&\\[-3mm]
 $G_0=H+\cdots$ & $\ell$ & $N$   \\[1mm]
\hline \hline &&\\[-3mm]
$sl(k|l)\oplus B(m-k|n-l)$ & 5 & $(k+l)(2m-2k+2n-2l+1)$  \\
         ($k=0,\ldots,m$; $l=0,\ldots,n$; & &  \\
          $(k,l)\not\in\{(0,0),(1,0)\} )$ && \\[1mm] 
         \cline{1-3} 
         &&\\[-3mm] 
         $B(m-1|n) \qquad [(k,l)=(1,0)]$  & 3 & $2m+2n-1$  \\[1mm] 
\hline\hline         
\end{tabular}
\vskip 0.5cm

\noindent
The most interesting case is with $k=m, l=n$. Then $G_0=sl(m|n)$,
$N=n+m$ and the CAOs:

\begin{eqnarray}
&&b_{j}^-\equiv B_{j}^-= -\sqrt{2}(e_{2m+1, 2m+1+n+j}+e_{2m+1+j,2m+1}), \nn\\
&&b_{j}^+\equiv B_{j}^+= \sqrt{2}(e_{2m+1, 2m+1+j}-e_{2m+1+n+j,2m+1}), \nn\\
&&b_{n+k}^-\equiv F_{k}^-= \sqrt{2}(e_{k, 2m+1}-e_{2m+1,m+k}), \nn \\
&& b_{n+k}^+\equiv  F_{k}^+=\sqrt{2}(e_{2m+1,k}-e_{m+k,2m+1}),\nn\\
&& \quad j=1,\ldots , n;\;\; k=1,\ldots,m, \nn
\end{eqnarray}
with
\[
\deg(b_j^\pm)=\langle j\rangle= \left\{ \begin{array}{lll}
 {1} & \hbox{if} & j=1,\ldots ,n \\ 
 {0} & \hbox{if} & j=n+1,\ldots ,n+m
 \end{array}\right.
\]
satisfy only triple relations:
\begin{eqnarray}
&& \lb\lb b_{ j}^{\xi}, b_{ k}^{\eta}\rb , b_{l}^{\epsilon}\rb =-2
\delta_{jl}\delta_{\epsilon, -\xi}\epsilon^{\langle l \rangle} 
(-1)^{\langle k \rangle \langle l \rangle }
b_{k}^{\eta} +2  \epsilon^{\langle l \rangle }
\delta_{kl}\delta_{\epsilon, -\eta}
b_{j}^{\xi},  \nn\\
&& \qquad\qquad \xi, \eta, \epsilon =\pm\hbox{ or }\pm 1;\quad j,k,l=1,\ldots,n+m. \nn 
\end{eqnarray}
Note that $B_j^\pm, j=1,\ldots, n$ (resp.\ $F_k^\pm, k=1,\ldots,m$) are para-Bose~(\ref{para-Bose})
(resp.\ para-Fermi~(\ref{para-Fermi})) CAOs.
The fact that $B(m|n)$ can be generated by $n$ pairs of para-Bose and $m$ pairs of
para-Fermi operators has been discovered by Palev~\cite{Posp}. 

In the next subsections we summarize the classification process for the Lie superalgebras 
$B(0|n)$, $D(m|n)$ and $C(n)$.

\subsection{The Lie superalgebras $B(0|n)$}

\bigskip
\noindent
{
\begin{tabular}{l|l|l}
\hline &&\\[-3mm]
 $G_0=H+\cdots$ & $\ell$ & $N$   \\[1mm] 
\hline \hline &&\\[-3mm]
$sl(i)\oplus B(0|n-i)$ & 5 & $i(2n-2i+1)$  \\ 
          ($i=1,\ldots,n$) & & \\[1mm]
\hline\hline         
\end{tabular}
}         

\vskip 1cm
\noindent
The most interesting case corresponds to $i=n$. Then  $N=n$; the CAOs
\begin{eqnarray}
&&B_{j}^-= -\sqrt{2}(e_{1, 1+n+j}+e_{1+j,1}), 
\quad j=1,\ldots , n,\nn\\
&&B_{j}^+= \sqrt{2}(e_{1, 1+j}-e_{1+n+j,1}), 
\quad j=1,\ldots , n\nn
\end{eqnarray}
are all odd generators of $B(0|n)$ and 
the relations ${\cal R}$ consists of the triple para-Bose relations~(\ref{para-Bose}).

\subsection{The Lie superalgebras $D(m|n)$}

\bigskip
\noindent
{
\begin{tabular}{l|l|l}
\hline &&\\[-3mm]
$G_0=H+\cdots$ & $\ell$ & $N$   \\[1mm] 
\hline \hline &&\\[-3mm] 
 $sl(k|l)\oplus D(m-k|n-l)$ & 5 & $2(k+l)(m+n-k-l)$   \\
         ($k=0,1,\ldots,m$;  & &  \\
          $l=0,1,\ldots,n$; && \\
          $(k,l)\not\in\{(0,0),(1,0),(m-1,n),(m,n)\} $) &&\\[1mm] 
         \cline{1-3} &&\\[-3mm]
         $D(m-1|n)\qquad [(k,l)=(1,0)]$  & 3 & $2(m+n-1)$  \\[1mm] 
         \cline{1-3} &&\\[-3mm] 
          $sl(m|n) \qquad [(k,l)=[m,n)]$ & 3 & $\frac{(m+n)(m+n+1)}{ 2}-m$  \\[1mm] 
         \cline{1-3}&&\\[-3mm] 
         $sl(m-1|n) \qquad [(k,l)=(m-1,n)]$ & 5 & $\frac{(m+n)(m+n+1)}{ 2}-m$  \\ 
         \cline{1-3} &&\\[-3mm]
         $sl(m-1|n)\qquad [(k,l)=(m-1,n)]$ & 5 & $2(m+n-1)$  \\[1mm] 
\hline\hline         
\end{tabular}
} 

\subsection{ The Lie superalgebras $C(n)$}

\bigskip
{
\begin{tabular}{l|l|l}
\hline &&\\[-3mm]
$G_0=H+\cdots$ & $\ell$ & $N$   \\[1mm]
\hline \hline &&\\[-3mm]
$sl(k|l)\oplus D(1-k|n-1-l)$ & 5 & $2(k+l)(n-k-l)$  \\
          ($k=0,1$; $l=1,\ldots,n-2$) &  \\[1mm]
         \cline{1-3} &&\\[-3mm]         
          $C_{n-1}\qquad [(k,l)=(1,0)]$ & 3 & $2(n-1)$  \\[1mm]
         \cline{1-3} &&\\[-3mm]
          $sl(1|n-1)\qquad [(k,l)=(1,n-1)]$ & 3 & $n(n+1)/2-1$  \\[1mm]
         \cline{1-3}&&\\[-3mm]
          $sl(n-1)\qquad [(k,l)=(0,n-1)]$ & 5 & $n(n+1)/2-1$ \\[1mm]
         \cline{1-3} &&\\[-3mm]
          $sl(n-1)\qquad [(k,l)=(0,n-1)]$ & 5 & $2(n-1)$  \\[1mm]
\hline\hline         
\end{tabular}
}

\bigskip

\section{Conclusions and possible applications}

We have obtained a complete classification of all GQS associated with 
the basic classical Lie superalgebras. The familiar cases (para-Bose, para-Fermi  and
$A$-(super)statistics) appear as simple examples in our classification.
In order to talk about a quantum statistics in the physical sense,
one should take into account additional requirements for the
CAOs, related to certain quantization postulates.
 These conditions are related to the existence of state spaces,
in which the CAOs act in such a way that the corresponding observables
are Hermitian operators.
We hope that some cases of our classification will yield interesting
GQS also from this point of view.

As a second application, we mension the problem of
finding solutions of the compatibility conditions (CCs)
of a Wigner quantum oscillator system~\cite{WQS}. These compatibility conditions
take the form of certain triple relations for operators.
So formally the CCs appear as special triple relations among
operators which resemble the creation and annihilation operators of
a generalized quantum statistics. One can thus investigate which
formal GQSs also provide solutions of the CCs.
It turns out that the classification presented here, with  CAOs
consisting of odd generators only, yields 
new solutions of these compatibility conditions corresponding to each 
basic classical Lie superalgebra~\cite{CC}.
%


\section*{Acknowledgments}
N.I.\ Stoilova was supported by a project from the Fund for Scientific Research, Flanders (Belgium).



\begin{thebibliography}{10}


\bibitem{Green}
H.S. Green, {\it Phys. Rev.} {\bf 90}, 270 (1953).

\bibitem{K}
S. Kamefuchi  and Y. Takahashi, 
{\it  Nucl.\ Phys.} {\bf 36}, 177 (1962).

\bibitem{S} 
C.\ Ryan and E.C.G.\ Sudarshan,
{\it Nucl.\ Phys. {\bf 47}, 207 (1963)}.

\bibitem{Ganchev}
A.Ch.\ Ganchev and T.D.\ Palev,  {\it J.\ Math.\ Phys.} {\bf 21}, 797 (1980).
 
\bibitem{Kac}
V.G.\ Kac, {\it Adv. Math.} {\bf 26}, 8 (1977).

\bibitem{Palev1}
T.D.\ Palev, {\it Lie algebraical aspects of the quantum statistics,} (Habilitation thesis,
Inst. Nucl. Research and Nucl. Energy, Sofia, 1976, in Bulgarian).

\bibitem{Palev2}
T.D.\ Palev, {\it Lie algebraic aspects of quantum statistics. Unitary quantization (A-quantization)},
Preprint JINR E17-10550 (1977) and hep-th/9705032.

\bibitem{Palev3}
T.D.\ Palev, {\it Czech.\ J.\ Phys.} {\bf B 29}, 91 (1979); 
T.D.\ Palev, {\it A-superquantization}, Communication JINR E2-11912 (1978).

\bibitem{Palev4}
T.D.\ Palev, {\it Rep.\ Math.\ Phys.} {\bf 18}, 117 (1980); {\bf 18}, 129 (1980).

\bibitem{Palev5}
T.D.\ Palev, {\it J.\ Math.\ Phys.} {\bf 21}, 1293 (1980).

\bibitem{PalevJeugt}
T.D.\ Palev, J.\ Van der Jeugt,
 {\it J.\ Math.\ Phys.} {\bf 43}, 3850 (2002).

\bibitem{Jellal}
A.\ Jellal, T.D.\ Palev and J.\ Van der Jeugt,
 {\it J.\ Phys.\ A: Math.\ Gen.} {\bf 34}, 10179 (2001); preprint
hep-th/0110276.

\bibitem{sl(m|n)}
T.D.\ Palev, N.I.\ Stoilova and J.\ Van der Jeugt, 
{\it J.\ Phys.\ A: Math.\ Gen.} {\bf 33}, 2545 (2000).

\bibitem{sl(1|n)}
T.D.\ Palev, N.I.\ Stoilova and J.\ Van der Jeugt, 
{\it J.\ Phys.\ A: Math.\ Gen.} {\bf 36}, 7093 (2003).

\bibitem{GQS}
N.I.\ Stoilova and J.\ Van der Jeugt,
{\it J.\ Math.\ Phys.} {\bf 46}, 033501 (2005).

\bibitem{CLS}
N.I.\ Stoilova and J.\ Van der Jeugt, 
{\it J.\ Math.\ Phys.} {\bf 46}, 113504 (2005).

\bibitem{Posp}
T.D.\ Palev, {\it J.\ Math.\ Phys.} {\bf 23}, 1100 (1982).

\bibitem{WQS}
T.D.\ Palev, {\it J.\ Math.\ Phys.} {\bf 23}, 1778 (1982);
T.D.\ Palev, {\it Czech. Journ. Phys.} {\bf B32}, 680 (1982);
A.H.\ Kamupingene, T.D.\ Palev and S.P.\ Tsaneva, {\it J.\ Math.\ Phys. }
{\bf 27}, 2067 (1986).

\bibitem{CC}
N.I.\ Stoilova and J.\ Van der Jeugt,
{\it J.\ Phys. A} {\bf 38}, 9681 (2005).
\end{thebibliography}
\end{document}